\documentclass[twocolumn, superscriptaddress,amsmath, prl, aps]{revtex4-2}

\usepackage[T1]{fontenc}
\usepackage[utf8]{inputenc}
\usepackage{siunitx}
\usepackage{times,color,amsthm,graphics,graphicx,bm,bbm,dcolumn}
\usepackage{epsfig}
\usepackage{graphicx}
\usepackage{xcolor}
\usepackage[colorlinks,urlcolor=black,citecolor=blue,linkcolor=black]{hyperref}
\usepackage{soul,bm}
\usepackage{verbatim}
\usepackage{orcidlink}
\setcitestyle{super}

\begin{document}
\setlength{\tabcolsep}{18pt}

\author{M.~Fr\'ometa Fern\'andez~\orcidlink{0000-0003-4937-8306}}
\altaffiliation{These authors contributed equally to this work}
\email[E-mail: ]{marcia.frometa@lens.unifi.it, rajkov@lens.unifi.it}
\affiliation{European Laboratory for Nonlinear Spectroscopy (LENS), University of Florence, 50019 Sesto Fiorentino, Italy}
\affiliation{Istituto Nazionale di Ottica del Consiglio Nazionale delle Ricerche (CNR-INO) c/o LENS, 50019 Sesto Fiorentino, Italy}
\affiliation{INFN, Sezione di Firenze, 50019 Sesto Fiorentino, Italy}

\author{D.~Hern\'andez-Rajkov~\orcidlink{0009-0002-1908-4227}}
\altaffiliation{These authors contributed equally to this work}
\email[E-mail: ]{marcia.frometa@lens.unifi.it, rajkov@lens.unifi.it}
\affiliation{European Laboratory for Nonlinear Spectroscopy (LENS), University of Florence, 50019 Sesto Fiorentino, Italy}
\affiliation{Istituto Nazionale di Ottica del Consiglio Nazionale delle Ricerche (CNR-INO) c/o LENS, 50019 Sesto Fiorentino, Italy}
\affiliation{INFN, Sezione di Firenze, 50019 Sesto Fiorentino, Italy}
\altaffiliation{These authors contributed equally to this work.}

\author{G.~Del Pace \orcidlink{0000-0002-0882-2143}}
\affiliation{European Laboratory for Nonlinear Spectroscopy (LENS), University of Florence, 50019 Sesto Fiorentino, Italy}
\affiliation{Istituto Nazionale di Ottica del Consiglio Nazionale delle Ricerche (CNR-INO) c/o LENS, 50019 Sesto Fiorentino, Italy}
\affiliation{INFN, Sezione di Firenze, 50019 Sesto Fiorentino, Italy}
\affiliation{Department of Physics, University of Florence, 50019 Sesto Fiorentino, Italy}

\author{N.~Grani~\orcidlink{0000-0001-6107-9726}}
\affiliation{European Laboratory for Nonlinear Spectroscopy (LENS), University of Florence, 50019 Sesto Fiorentino, Italy}
\affiliation{Istituto Nazionale di Ottica del Consiglio Nazionale delle Ricerche (CNR-INO) c/o LENS, 50019 Sesto Fiorentino, Italy}
\affiliation{INFN, Sezione di Firenze, 50019 Sesto Fiorentino, Italy}
\affiliation{Department of Physics, University of Florence, 50019 Sesto Fiorentino, Italy}

\author{M.~Inguscio \orcidlink{0000-0001-8152-8103}}
\affiliation{European Laboratory for Nonlinear Spectroscopy (LENS), University of Florence, 50019 Sesto Fiorentino, Italy}
\affiliation{Istituto Nazionale di Ottica del Consiglio Nazionale delle Ricerche (CNR-INO) c/o LENS, 50019 Sesto Fiorentino, Italy}

\author{F.~Scazza~\orcidlink{0000-0001-5527-1068}}
\affiliation{European Laboratory for Nonlinear Spectroscopy (LENS), University of Florence, 50019 Sesto Fiorentino, Italy}
\affiliation{Department of Physics, University of Trieste, 34127 Trieste, Italy}
\affiliation{Istituto Nazionale di Ottica del Consiglio Nazionale delle Ricerche (CNR-INO), 34149 Trieste, Italy}

\author{S.~Stringari \orcidlink{0000-0002-5960-0612}}
\affiliation{Pitaevskii BEC Center, CNR-INO and Dipartimento di Fisica, Universit\`a di Trento, Via Sommarive 14, 38123 Povo, Trento, Italy}
\affiliation{Trento Institute for Fundamental Physics and Applications, INFN, 38123 Povo, Italy}

\author{G.~Roati \orcidlink{0000-0001-8749-5621}}
\affiliation{European Laboratory for Nonlinear Spectroscopy (LENS), University of Florence, 50019 Sesto Fiorentino, Italy}
\affiliation{Istituto Nazionale di Ottica del Consiglio Nazionale delle Ricerche (CNR-INO) c/o LENS, 50019 Sesto Fiorentino, Italy}
\affiliation{INFN, Sezione di Firenze, 50019 Sesto Fiorentino, Italy}

\title{Angular momentum of rotating fermionic superfluids by Sagnac phonon interferometry}

\date{\today}

\begin{abstract}
Fermionic many-body systems provide an unrivaled arena to investigate how interactions drive the emergence of collective quantum behavior, such as macroscopic coherence and superfluidity~\cite{leggett2006quantum}. Central to these phenomena is the formation of Cooper pairs, correlated states of two fermions that behave as composite bosons and condense below a critical temperature. However, unlike elementary bosons, these pairs retain their internal structure set by underlying fermionic correlations, essential for understanding superfluid properties throughout the so-called Bose–Einstein condensate (BEC) to Bardeen–Cooper–Schrieffer (BCS) crossover --- a cornerstone of strongly correlated fermionic matter~\cite{zwerger2012bcs}. Here, we harness a sonic analog of the optical Sagnac effect~\cite{sagnac1913effet} to disclose the composite nature of fermionic condensates across the BEC–BCS crossover. We realize an in-situ loop interferometer by coherently exciting two counter-propagating long-wavelength phonons of an annular fermionic superfluid with tuneable interparticle interactions. The frequency degeneracy between clock- and anticlock-wise sound modes is lifted upon controllably injecting a quantized supercurrent in the superfluid ring~\cite{DelPace2022}, resulting in a measurable Doppler shift that enables us to probe the elementary quantum of circulation and the angular momentum carried by each particle in the fermionic fluid~\cite{Kumar2016, Woffinden2023autralians}.  Our observations directly reveal that the superflow circulation is quantized in terms of $h/2m$, where $m$ is the mass of the constituents, in striking contrast to bosonic condensates where $h/m$ is the relevant circulation quantum.  Further, by operating our interferometer at tunable temperature, we measure the thermal depletion of the superfluid in the unitary Fermi gas, demonstrating phonon interferometry as a powerful technique for probing fundamental properties of strongly-correlated quantum systems.
\end{abstract}

\maketitle
\normalsize

Rotating a quantum system is amongst the most effective strategies for both uncovering its macroscopic behavior and linking it to microscopic features. A hallmark of superfluidity, as anticipated by Onsager~\cite{Onsager1949} and Feynman~\cite{Feynman1955}, is the quantization of circulation that emerges from the single-valuedness of the macroscopic wavefunction describing the entire quantum system (see Ref.[~\citenum{Zieve2023}] for a recent review). Superfluid flow is constrained by the discrete quantization of its circulation $\oint {\bf v}_s \cdot \mathrm{d}{\bf l}$ into units of $\kappa$, i.e.~the quantum of circulation, which provides insight into the very mechanisms driving the emergence of superfluidity. Pioneering measurements of $\kappa$ in superfluid $^3$He-B~\cite{Davis1991, Simola1989, Packard1992} yielded $h/2m$ (where $h$ is Planck’s constant and $m$ mass of a single atom), contrasting with He-II result of $h/m$~\cite{Vinen1956, Zimmermann1968, Hess1967} and thereby confirming that transport in $^3$He occurs through Cooper pairs of opposite spin rather than single particles~\cite{vollhardt2013superfluid}. Similarly, for superconductors the single-valuedness of the electronic wavefunction enforces flux quantization, $\Phi_0 = \oint {\bf A}\!\cdot \!\mathrm{d}{\bf l}$, with the vector potential ${\bf A}$ playing the role of the superfluid velocity in the case of charged particles. Early observations in superconducting rings confirmed charge-$2e$ Cooper pairs~\cite{Fairbank1961, Näbauer1961} for standard superconductors, and recent studies suggest charge-$4e$ and $6e$ states~\cite{ge2024charge} for more exotic ones. In ultracold atomic Fermi gases, the formation of Abrikosov vortex lattices in rotating traps offered the first striking evidence of superfluidity across the BEC–BCS crossover~\cite{Zwierlein2005}. However, only indirect estimates of $\kappa$ could be extracted there from the inter-vortex spacing at a given rotation velocity. Complementary insights came from angular momentum measurements in slowly rotating Fermi gases, which revealed deviations of the moment of inertia from the classical rigid-body value highlighting the role of temperature~\cite{Riedl2009, Riedl2011}.

\begin{figure}[t!]
\centering
\includegraphics[width=8.5cm]{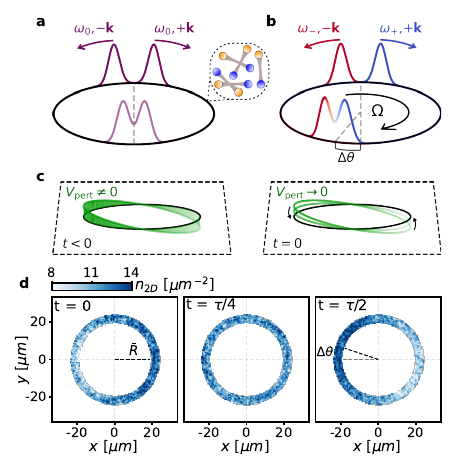}
\caption{\textbf{Sagnac-like matter-wave interferometer with annular Fermi superfluids.}
In analogy with optical Sagnac interferometers utilizing counter-propagating light waves in a loop, our interferometer is based on the excitation of two counter-propagating sound modes in an annular fermionic-pair condensate with tunable interactions. In the absence of rotation (\textbf{a}), both excitations follow the dispersion relation $\omega_0 = c_s k$. In a rotating system (\textbf{b}), they experience a relative Doppler shift, yielding $\omega_{\pm} = c_s k \pm \Omega$, where the frequency splitting $\Omega$ is proportional to the angular momentum per particle. Such Doppler splitting manifests as a precession of the standing-wave atomic density pattern, accruing a phase shift $\Delta \theta = \Omega t$ over a time $t$. (\textbf{c}), In our phonon loop-interferometer, the longest-wavelength phonon modes with $\textbf{k}\bar{R} = \pm \hat \theta$ are simultaneously excited by a weak potential perturbation $V_\text{pert} = -V_0 \cos{\theta}$ that is suddenly quenched off at $t=0$, defining the onset of the dynamics. (\textbf{d}), \textit{In-situ} atomic density profiles recorded at different evolution times reveal the precession of the standing wave, with the angular shift $\Delta\theta = \Omega t$, from which we experimentally extract $\Omega$. Here, $\tau = 2\pi/\omega_0$ denotes the phonon oscillation period.
}
\label{fig1}
\end{figure}

In this work, we realize a Sagnac-like phonon interferometer to measure the quantum of circulation $\kappa$ in annular atomic Fermi superfluids across the BEC–BCS crossover. In analogy to optical Sagnac gyroscopes that detect rotation-induced phase shifts between counter-propagating light beams, we exploit a rotation-induced phononic phase shift to precisely quantify the angular momentum per particle. We find $\kappa$ to equal $h/2m$ within experimental uncertainties, both in the BEC regime of tightly bound dimers as well as in the strongly interacting unitary Fermi gas of extended pairs. We thus realize the first direct, \textit{in-situ} rotation sensing protocol for fermionic quantum fluids, complementing previous measurements in rotating bosonic condensates~\cite{chevy2000measurement}, where $\kappa$ was found to equal $h/m$. Leveraging the periodic boundary conditions of an annular trapping geometry, we excite two long-lived counter-propagating phonon modes $(\omega_\pm, \pm\textbf{k})$, whose interference creates a circular standing-wave density pattern with periodicity $2\pi/k$. Circulating currents alter the propagation velocity $c_s$ of these modes owing to a Doppler shift $\Delta_c$, effectively modifying the sonic dispersion relation as $\omega_{\pm} = c_s k \pm \Delta_c k$. This frequency splitting leads to a precession of the standing wave at angular frequency $\Omega = \Delta_c k$, such that the enclosed Sagnac phase appears as the angular shift $\Delta \theta = \Omega t$ (see Fig.~\ref{fig1}a-b).

At zero temperature, the Doppler shift $\Delta_c$ coincides with the background superfluid velocity $v_s$, as expected from a Galilean transformation that brings the system into the rest frame of the superfluid. The identity $\Delta_c = v_s$ has profound implications, as $v_s$ is quantized and proportional to $\kappa$, according to $v_s = \textit{w} \kappa / (2\pi \bar R)$, where $\textit{w} = 0, \pm1, \pm2, \dots$ denotes the integer winding number of the persistent current, and $\bar R$ the mean ring trap radius. Consequently, measuring $\Delta_c$ would allow to directly probe that: (i) the quantum of circulation for a Fermi superfluid is $\kappa=h/2m$ regardless of the interaction regime, and (ii) the angular momentum per particle $\ell_z = m \bar Rv_s$ is quantized according to the law $\ell_z= m \textit{w} \kappa/2\pi$.

At finite temperature, Landau's two-fluid hydrodynamic theory \cite{Landau1941} predicts the propagation of two distinct sound modes, known as first and second sound. When the velocities of the normal and superfluid components are different, the Doppler shifts of the two sounds exhibit a more complex dependence~\cite{nepomnyashchy1995extraordinary, kenis1999unusual, nepomnyashchy1993unusual, Tomasz2025}. As first pointed out by Khalatnikov \cite{Khalatnikov1956}, in a weakly compressible fluid such as liquid $^4$He, the Doppler shift of first sound—being a density wave—is fixed by the fluid flow according to the relation $\Delta_c = f_s v_s + f_n v_n$, where $v_{s,n}$ are the velocities of the superfluid and normal components, and $f_{s,n} = N_{s,n}/N$ are their respective fractions, with $f_s + f_n = 1$. Note that only if $v_n=v_s$ the Doppler shift reduces to the velocity of the fluid. In the same limit of weakly compressible fluids, the Doppler effect of second sound exhibits a different anomalous behavior \cite{Khalatnikov1956}. The above relation for $\Delta_c$ is expected to hold also in superfluid Fermi gases near unitarity (see Methods). An interesting situation arises when the circulating current is carried solely by the superfluid component, i.e., when $v_n=0$. In this case, the measurement of the Doppler shift $\Delta_c$ provides access to $\ell_z$ according to the relation $\ell_z=m \bar R\Delta_c$, taking the form
\begin{equation}
\label{eq:lz}
\ell_z = \frac{m \textit{w}}{2\pi}\, \frac{N_s}{N}\; \kappa,
\end{equation}
which reduces to the $T=0$ case for $N=N_s$. The result in Eq.~(\ref{eq:lz}) opens the interesting possibility of measuring the superfluid density at finite temperature, as we will discuss later.

\begin{figure*}[ht!]
\centering
\includegraphics[height=8cm]{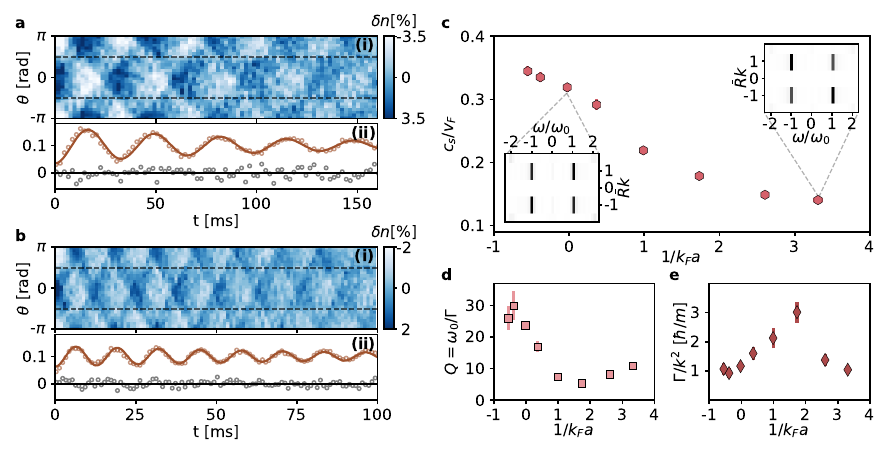}
\caption{\textbf{Phonon interferometry across the BEC–BCS crossover.} Temporal evolution of the azimuthal density modulation $\delta n(\theta,t)$ for a BEC at $1/k_Fa = 3.31(7)$ (\textbf{a}) and a unitary Fermi gas (\textbf{b}), both in the absence of circulation. Dashed lines in panels \textbf{a}(i) and \textbf{b}(i) are a guide to the eye of the nodal points of the standing wave. Panels \textbf{a}(ii) and \textbf{b}(ii) display the corresponding density-weighted angular moments $\langle \cos\theta \rangle$ (brown) and $\langle \sin\theta \rangle$ (black), as defined in the main text. For visual clarity, $\langle \cos\theta \rangle$ is vertically offset by $+0.1$ and $\langle \sin\theta \rangle$ is multiplied by 3. To extract the phonon frequency $\omega_0$, we fit the density modulations $\delta n(\theta,t)$ using Eq.~\eqref{eqlinearresponse}; the resulting fits are shown as continuous lines on the moments traces. (\textbf{c}), The phonon velocity, expressed as $c_s/v_F = \omega_0 \bar{R}/v_F$ with $v_F = \sqrt{2mE_F}$ the Fermi velocity. Insets show the corresponding power spectral densities of panels \textbf{a}(i) and \textbf{b}(i), confirming the monochromatic excitation with $\bar{R}k = \pm1$, and the absence of second-sound coupling. (\textbf{d}), Quality factor $Q = \omega_0/\Gamma$ across the BEC–BCS crossover. (\textbf{e}), Damping rate $\Gamma/k^2$ expressed in units of $\hbar/m$. Error bars in panels \textbf{c}-\textbf{e} stem from the standard fitting error of $\delta n$.
}
\label{fig2}
\end{figure*}

We produce annular Fermi superfluids by cooling a balanced mixture of the first and third lowest hyperfine states of $\mathrm{^6Li}$ atoms, with $N_{\rm p} = \SI{6.5\pm 0.1}{} \times 10^3$ atoms per spin component. The interaction strength is parametrized by $1/k_F a$, where $k_F = \sqrt{2 m E_F}/\hbar$ is the Fermi wave vector corresponding to the Fermi energy $E_F$, and $a$ is the $s$-wave scattering length tuned via a Feshbach resonance at $\SI{690}{G}$. This allows to explore the BEC–BCS crossover from a molecular BEC (\(1/k_F a > 1\)) to a weakly paired BCS state (\(1/k_F a < 0\)), including the unitary regime (\(1/k_F a \to 0\)). The gas is confined in the $x$–$y$ plane by an annular-shaped hard-wall potential, creating a ring with typical radial extent of $\bar R\simeq22\,\mu$m, and width $\simeq7\,\mu$m (see Methods). Along the vertical $z$-direction, the atoms experience a tight harmonic confinement with trapping frequency $\omega_z \simeq 2\pi\times \SI{604}{Hz}$. This geometry produces an in-plane density that is nearly homogeneous, while keeping the system in a three-dimensional thermodynamic regime across all interaction strengths. Measurements are performed at temperatures $T$ between $0.07\,T_F$ and $0.12\,T_F$, where $T_F = E_F/k_B$ is the Fermi temperature, and $k_B$ the Boltzmann constant, ensuring that the system remains below the superfluid transition temperature $T_c$~\cite{NGrani2025MutualFriction}, throughout the explored interaction range. 

The interferometric sequence begins by raising a weak perturbation potential $V_{\text{pert}} =- V_0 \cos(\theta)$ (see Fig.~\ref{fig1}c). The perturbation strength $V_0$ is kept small to ensure operation within the linear-response regime (see Methods). We allow the system to equilibrate for 50 ms, damping the residual motion of the normal component \cite{Riedl2009} along with other spurious excitations. The perturbation is then suddenly turned off, generating two counter-propagating monochromatic excitations with the lowest available wavenumber in the system, $|\mathbf{k}|=1/\bar{R}$ (see Fig.~\ref{fig1}d). We note that similar phonon interferometric schemes have been explored in annular bosonic superfluids~\cite{Kumar2016, Sagnac_Marti2015, Woffinden2023autralians}.

\begin{figure*}[ht!]
\centering
\includegraphics[height=8cm]{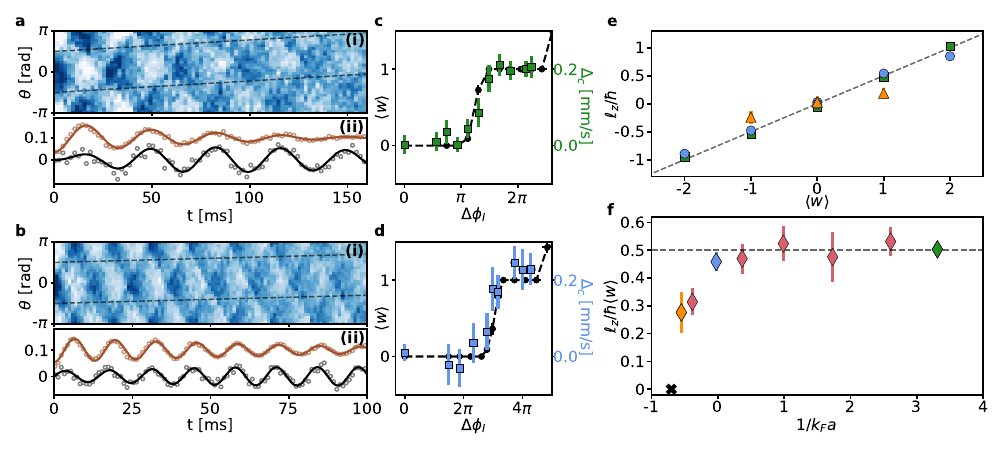}
\caption{
\textbf{Phonon Doppler shift and angular momentum per particle across the BEC–BCS crossover.} 
Temporal evolution of the azimuthal density modulation $\delta n(\theta,t)$ in a rotating superfluid with a single quantum of circulation ($\textit{w} = 1$) for a BEC at $1/k_Fa = 3.31(7)$ (\textbf{a}) and a unitary Fermi gas (\textbf{b}). The color-scales are the same of Fig.~\ref{fig2}a-b, respectively. Dashed lines in panels \textbf{a}(i) and \textbf{b}(i) are guides to the eye of the nodal points of the standing wave. Panels \textbf{a}(ii) and \textbf{b}(ii) display the corresponding density-weighted angular moments $\langle \cos\theta \rangle$ (brown) and $\langle \sin\theta \rangle$ (black); for clarity, $\langle \cos\theta \rangle$ is offset by $+0.1$ and $\langle \sin\theta \rangle$ is multiplied by 3. The oscillatory behavior of $\langle \sin\theta \rangle(t)$ associated to the precession of the density modulation provides an unambiguous signature of rotation. We extract the Doppler shift $\Delta_c$, fitting $\delta n(\theta,t)$ with Eq.~\eqref{eqlinearresponse}, as a function of the total imprinted phase employed to generate the persistent current for the BEC (\textbf{c}), and UFG (\textbf{d}) regimes. In addition, we show the mean winding number $\langle\textit{w}\rangle$ measured over repeated acquisitions after the imprinting protocol. (\textbf{e}), Measured angular momentum per particle $\ell_z=m\bar R\Delta_c$ as a function of winding number $\textit{w}$ for a BEC at $1/k_Fa = 3.31(7)$ (green squares), a unitary Fermi gas (blue circles), and a BCS superfluid at $1/k_Fa = -0.55(1)$ (orange triangles). (\textbf{f}), Angular momentum per particle across the BEC–BCS crossover. The black cross marks the critical interaction strength below which persistent currents are no longer observed. Colors of symbols identify measurements in panel \textbf{e}. The dashed line marks $\ell_z / \hbar\textit{w}= 1 / 2$. Error bars in panels \textbf{c}-\textbf{d} stem from the standard fitting error of $\delta n$, whereas in panels \textbf{e}-\textbf{f} they represent the standard error of the mean over at least 3 repetitions.
}
\label{fig3}
\end{figure*}

We first test the interferometer response for non-rotating systems. For each interaction regime throughout the crossover, the time-evolution of the interference pattern is tracked by imaging the \textit{in-situ} atomic density. From these measurements, we extract the normalized azimuthal density profile defined as $n_{\theta}(\theta) = \int n_\text{3D}(r,\theta, z)\, r\, dr\, dz / N$. Figure~\ref{fig2}a-b shows the temporal evolution of the density modulation, $\delta n(\theta, t) = n_{\theta}(\theta, t) - n_{\theta}^0$, where the equilibrium profile $n_{\theta}^0$ is determined from measurements of the unperturbed cloud. At finite temperature, density excitations generally couple to both first and second sound modes, resulting in two characteristic modulation frequencies for each wavenumber $k$. The relative contribution of each mode to the density is governed by the density response function $\chi(k,\omega)$. Recent characterizations of $\chi(k,\omega)$ at unitarity indicate that for temperatures below $0.7\,T_c$, second sound couples only weakly to density~\cite{Li2022}, manifesting mostly as a entropy wave \cite{yan2024thermography}. In the temperature range explored here, our excitation protocol selectively excites only two counter-propagating first sound modes. This is confirmed by the power spectral density of the measured signal, which displays the peaks of the clockwise $(\pm\omega_0=\pm c_s \bar R^{-1})$ and anti-clockwise $(\pm\omega_0=\mp c_s \bar R^{-1})$ phonon modes (see insets of Fig.~\ref{fig2}c), without populating additional peaks -- e.g.~at frequencies $|\omega| \approx 0.26\,\omega_0$, expected for second sound excitations~\cite{yan2024thermography, Li2022}. In the absence of a circulating current, the density modulations resulting from quenching off the static perturbation $V_{\text{pert}}$ takes the form $\delta n(\theta,t)=e^{-\Gamma t/2}\,V_0 (\chi/2) \cos(\theta)\cos(\omega_0t)$, with a time dependence fixed by the first sound frequency $\omega_0$ and the damping coefficient $\Gamma$, which sets the decay of the collective oscillations. Here, $\chi$ is the static response that coincides with the isothermal compressibility $(\partial \rho/\partial p)_T$ according to the compressibility sum rule \cite{PitaevskiiStringariBook2016}. Furthermore, we analyze the density-weighted angular moments $\langle \cos{\theta} \rangle (t) = \int_0^{2\pi} n_{\theta}(\theta, t) \cos \theta\, d\theta$ and $\langle \sin{\theta} \rangle (t) = \int_0^{2\pi} n_{\theta}(\theta, t) \sin \theta\, d\theta$. In these observables, the presence of circulating currents is directly revealed by $\langle \sin{\theta} \rangle (t) \neq 0$. As shown in Fig.~\ref{fig2}a-b, this is consistently vanishingly small in the absence of superflow around the ring. By using the above relations, we fit the evolution of $n_{\theta}(\theta, t)$ to extract the phonon propagation speed $c_s$ and the corresponding $\Gamma$. The extracted values of $c_s$, shown in Fig.~\ref{fig2}c, increase monotonically from the BEC to the BCS regime, reflecting the rise in chemical potential through the crossover. Nevertheless, they are consistently lower than those found in homogeneous systems~\cite{biss2022excitation}, primarily due to the vertical density inhomogeneity of our trapped gases~\cite{joseph2007measurement}. At unitarity, the sound velocity is directly connected to the Bertsch parameter $\xi_B$, through its link to the thermodynamic compressibility~\cite{PitaevskiiStringariBook2016}. At the lowest temperature explored in this work, ($T/T_c \sim 0.4$), we measure $\xi_B = \SI{0.396\pm0.029}{}$ (see Methods). This value is consistent with previous determinations from thermodynamic measurements~\cite{ku2012revealing, navon2010equation}, sound-excitation experiments~\cite{Li2022, hoinka2017goldstone}, and quantum Monte Carlo calculations~\cite{carlson2011auxiliary, haussmann2008thermodynamics}.

The interferograms display high contrast over several phonon rotation periods, reflecting the long-lived character of long-wavelength sound modes in the ring, especially in the strongly interacting regime. To characterize this behavior more quantitatively, we extract the interferometer quality factor $Q=\omega_0/\Gamma$, from the phonon damping rate. The results, shown in Fig.~\ref{fig2}d for each interaction strength, reveal an enhancement of $Q\approx30$ for fermionic superfluids near unitarity. We can link this behavior to sound transport properties in strongly interacting systems. In the hydrodynamic collisional regime, the sound diffusivity $D=\Gamma/k^2$, approaches the fundamental limit fixed by the ratio $\hbar/m$, signaling the onset of \textit{quantum-limited} transport~\cite{hu2018low, zhang2011finite}. Rescaling $\Gamma$ by $k^2$ yields the behavior shown in Fig.~\ref{fig2}(e), which is consistent with previous results~\cite{yan2024thermography, Patel2020, martirosyan2024universal, sommer2011universal, bohlen2020sound}. As interactions are tuned away from unitarity towards the BEC side, $\Gamma/k^2$ first increases with the mean free path up to $1/k_Fa \sim 2$, then decreases again, likely reflecting the onset of a dimensional crossover (see Methods). While the microscopic mechanisms governing phonon damping across the crossover and their sensitivity to the effective dimensionality remain open challenges for future studies, our results establish strongly interacting Fermi superfluids as a promising platform for atom interferometry and sensing~\cite{Woffinden2023autralians}.

To access the angular momentum per particle $\ell_z$, we set the superfluid gas into rotation by using a phase-imprinting technique~\cite{DelPace2022}, known to provide enhanced stability and control compared with conventional stirring methods \cite{abo2001observation, chevy2000measurement, Zwierlein2005}. We inject quantized circulation of winding number $\textit{w}$ by optically imprinting a phase $\Delta \phi_{I}$ on the ring. In the presence of the persistent current, we excite the two phonon branches and study the Doppler effect on this rotating system. Unlike the non-rotating case, the density modulation $\delta n(\theta,t)$ exhibits a clear spatio-temporal drift, signaling a precession of the interference pattern (see Fig.~\ref{fig3}a-b for $\textit{w}=1$). Moreover, $\langle \sin \theta(t)\rangle$ shows clear oscillations with non-zero amplitude, revealing the presence of the background circulating flow. To quantify the precession rate, we fit the observed dynamics with the theoretical prediction of linear response theory:
\begin{align}\label{eqlinearresponse}
    &\delta n (\theta, t) = \frac{V_0}{2} e^{- \frac{\Gamma t}{2}} \left[\chi_+\cos(\theta-\omega_+t)+\chi_-\cos(\theta+\omega_-t)\right].
\end{align}
This expression explicitly accounts for the different time-dependent evolution of the two Doppler shifted components of the first sound mode, resulting in the precession $d\theta/dt = \Omega= \Delta_c/\bar R $ visible as a shift of the local zeros of $\delta n (\theta, t) $ (see guidelines in Fig.~\ref{fig3}a-b). In the above equation, $\chi_\pm=(\chi/2)(1\mp\Delta_c/c_s)$ are the corresponding contributions to the static response $ \chi_+ + \chi_-=\chi_1\simeq \chi$, neglecting the contribution of second sound to the response. 

In our system, the background rotational flow is expected to be sustained solely by the superfluid, as any residual incoherent motion of the normal component is damped during the equilibration period preceding the quench of $V_{\text{pert}}$, effectively yielding $f_n v_n \simeq 0$ when the potential is quenched. To corroborate that the contribution of normal fluid motion is negligible, we measured the Doppler shift $\Delta_c$ as a function of the imprinted phase~\cite{DelPace2022} $\Delta \phi_{I}$ (see Fig.~\ref{fig3}c-d) and found that, mirroring the measured mean winding number $\langle \textit{w} \rangle$, it exhibits a step-like behavior rather than the linear monotonous increase expected for a normal moment of inertia~\cite{Riedl2011, Riedl2009}. The visible deviation between the phase imprinted to obtain $\langle \textit{w} \rangle =1$ in a BEC or in a UFG is inherent to the optical phase-imprinting technique in fermionic gases \cite{DelPace2022, Chen2025pn}. For the following discussions, interferometric measurements were performed using the central value of each $\Delta \phi_{I}$ step to establish a well-defined circulation state, $\textit{w}$. Fig.~\ref{fig3}e shows the angular momentum per particle as a function of the winding number $\textit{w}$, for different interaction regimes. As expected, $\ell_z$ increases linearly with $\textit{w}$, or equivalently with $v_s$, with a slope close to the zero-temperature value $\hbar/2$. A deviation is observed in the BCS regime, as discussed next. In Fig.~\ref{fig3}f, we present the slope $\ell_z/\textit{w}$ as a function of $1/k_Fa$. In the BEC regime, $\ell_z/\textit{w} \approx \hbar/2$, consistent with superfluidity arising from tightly bound molecules. A comparable value is found at unitarity, indicating that the superflow is carried by fermion pairs rather than individual particles. In the BCS regime, however, $\ell_z/\textit{w}$ is strongly suppressed, which we attribute to the marked reduction of the superfluid fraction as $T_c/T_F$ rapidly decreases in this limit \cite{pisani2018entanglement}. These results are consistent with the critical value of $1/k_Fa$ beyond which the phase-imprinting protocol fails to generate persistent currents, $1/k_Fa \lesssim \SI{-0.7 \pm 0.1}{}$, indicated by a cross in Fig.~\ref{fig3}f, in agreement with the observed trend of $\ell_z$.

\begin{figure}[t!]
\centering
\includegraphics[width=8cm]{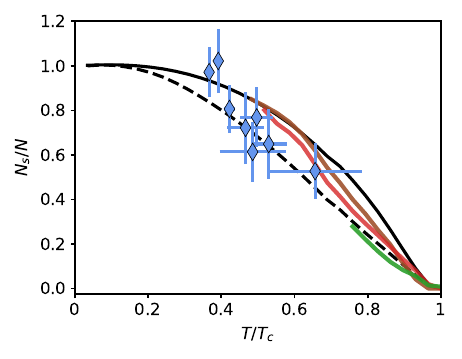}
\caption{\textbf{Superfluid fraction at unitarity.} We estimate the superfluid fraction extracted from the angular momentum per particle, obtained from the relation $N_s/N = 2m\bar R \Delta_c/\hbar$, as a function of the reduced temperature $T/T_c$, where $T_c\simeq0.18\,T_F$ in our trap~\cite{NGrani2025MutualFriction}. Here, we used the fact that the quantum of circulation at unitarity is $\kappa=h/2m$.  Color solid lines represent previous experimental extractions of the superfluid fraction in homogeneous unitary Fermi gases, adjusted to account for harmonic confinement along the $z$-axis:~Ref.[~\citenum{Li2022}] (green), Ref.[~\citenum{yan2024thermography}] (brown), and Ref.[~\citenum{sidorenkov2013second}] (red). Similarly, we compare our results with predictions~\cite{pisani2023inclusion} from mLPDA mean-field description (black dashed line) and extended GMB approach (black continuous line). Error bars represent the standard fitting error of $\delta n$, and $T/T_c$.
}
\label{fig4}
\end{figure}

Owing to the near-vanishing of $v_n \simeq 0$, as anticipated in Eq.~\eqref{eq:lz}, the gas superfluid fraction at finite temperature can be quantitatively inferred from the temperature dependence of $\ell_z$. Motivated by this, we performed measurements of $\ell_z$ at unitarity as a function of temperature in the range $0.4-0.6~T_c$. The results, shown in Fig.~\ref{fig4}, reveal a decrease in the superfluid fraction as the temperature increases. However, similarly to the measurements of $c_s/v_F$, the obtained values are lower than those reported for homogeneous systems \cite{sidorenkov2013second, yan2024thermography, Li2022}, likely due to the vertical density inhomogeneity. For a meaningful comparison, we extrapolate the superfluid fraction of Refs.[~\citenum{sidorenkov2013second, yan2024thermography, Li2022}] accounting for the effect of harmonic confinement along the $z$-axis through the equation of state at unitarity~\cite{ku2012revealing}, and the local density approximation (see Methods). Applying the same considerations, in Fig.~\ref{fig4} we compare also with the theoretical predictions~\cite{pisani2023inclusion} based on the extended Gorkov–Melik–Barkhudarov (GBM) approach to the Fermi superfluid phase, and the modified Local Phase Density Approximation (mLPDA) mean-field approach. Taking the vertical harmonic confinement into account brings our results into good agreement with both experiment and theory, confirming that the reduction of $\ell_z$ at finite temperature at unitarity and in the BCS regime originates from finite temperature effects which diminish the superfluid fraction. Our findings provide a novel benchmark for the low-temperature properties of paired superfluids and establish the factor-of-two reduction in the quantum of circulation, $\kappa=h/2m$, as a general hallmark of Fermi superfluidity. We stress that our interferometric technique measures the angular momentum per particle without relying on the precise knowledge of the system's equation of state; it could thus be readily extended to diverse systems, ranging from two-dimensional~\cite{Sobirey2021}, disordered~\cite{Geier2025} or periodically modulated~\cite{Tao2023,Chauveau2023} superfluids to dipolar supersolids~\cite{Sindik2024,Tomasz2025, Preti2025}.

A further direction opened by our work is the investigation of the anomalous Doppler effect of second sound~\cite{Khalatnikov1956, kenis1999unusual, nepomnyashchy1993unusual, Tomasz2025} via thermographic techniques~\cite{yan2024thermography}, thus allowing the study of heat and entropy transport in moving superfluids. To enable more direct comparisons with theoretical predictions and other experiments, it would be beneficial to employ 3D box-like trapping potentials that yield fully homogeneous atomic densities~\cite{navon2021quantum}. The ability to generate sound excitations with distinct wavevectors using tailored optical perturbation potentials will allow also for the systematic study of sound damping mechanisms, such as phonon–phonon and phonon–single-particle interactions, across the BEC–BCS crossover. This approach offers a direct and effective route to investigate how the curvature of the phonon dispersion affects energy dissipation and its connection to the onset of wave turbulence in strongly interacting Fermi superfluids~\cite{biss2022excitation, kurkjian2017three, Nazarenko2015, navon2016emergence}.

\section*{Acknowledgements}
We thank the Quantum Gases group at LENS for stimulating discussions.
We also thank S. Giorgini, F. Dalfovo, A. Recati, M. Šindik and T. Zawiślak and G. Strinati for valuable comments about this work. 
We thank the European Center for Theoretical Studies in Nuclear Physics and Related Areas (ECT*) in Trento for support at the Workshop "Nonequilibrium phenomena in superfluid systems: atomic nuclei, liquid helium, ultracold gases, and neutron stars", favoring fruitful discussions with M. Zwierlein.
G.R. and G.D.P. acknowledge financial support from the PNRR MUR project PE0000023-NQSTI. 
G.R. acknowledges funding from the Italian Ministry of University and Research under the PRIN2017 project CEnTraL and project CNR-FOE-LENS-2024. 
S.S. acknowledges support from the Provincia Autonoma di Trento.
F.S. acknowledges support from the EU under the Horizon 2020 research and innovation program (project OrbiDynaMIQs, GA No.~949438).
The authors acknowledge support from the European Union - NextGenerationEU within the “Integrated Infrastructure Initiative in Photonics and Quantum Sciences" (I-PHOQS). 
The authors acknowledge funding from INFN through the RELAQS project. 
This publication has received funding under the Horizon Europe programme HORIZON-CL4-2022-QUANTUM-02-SGA (project PASQuanS2.1, GA no.~101113690) and Horizon 2020 research and innovation programme (GA no.~871124).

\section*{Data availability}
The data supporting the figures within this paper will be available in a Zenodo repository.

\section*{Author contributions}
All authors contributed to the interpretation of the results and to the writing of the manuscript.

\section*{Competing interests} 
The authors declare no competing interests.

%

\clearpage
\section*{Methods}
\renewcommand{\figurename}{Extended Data Fig.}
\setcounter{figure}{0}

\subsection*{Superfluid ring preparation}
\begin{figure}[t!]
\centering
\includegraphics[width=8.6cm]{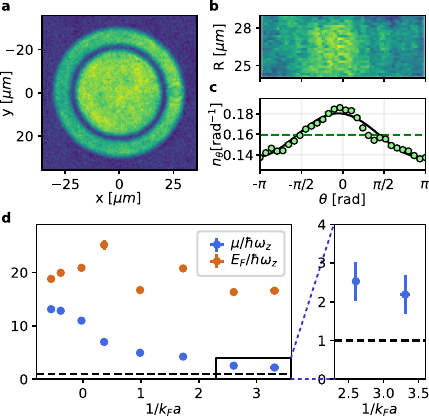}
\caption{\textbf{Density profiles and chemical potential.}
(\textbf{a}), In situ image of our annular superfluid in the unitary regime in presence of the perturbation potential $V_{\text{pert}}$. (\textbf{b}), To obtain the interferograms of Figs. \ref{fig2}-\ref{fig3}, we unwrap the angular coordinate, and (\textbf{c}), integrate along the radial direction to obtain a one-dimensional signal. The green dashed line represent the average homogeneous density profile $n_{\theta}^0$, which is $\simeq1/2\pi$. (\textbf{d}), Fermi energy and chemical potential to trap frequency ratio as a function of the interaction parameter.
}
\label{extfigMu}
\end{figure}

We prepare the unitary superfluid by evaporating a balanced mixture of the hyperfine states $|1\rangle=|F,m_F\rangle= |1/2,1/2\rangle$ and $|3\rangle=|F,m_F\rangle= |3/2,-3/2\rangle$ of $^6$Li, near their scattering Feshbach resonance at $\SI{690}{G}$ \cite{zurn2013precise} in an elongated, elliptic optical dipole trap. A repulsive $\mathrm{TEM_{01}}$-like optical potential at $\SI{532}{nm}$ is then adiabatically ramped up in $\SI{100}{ms}$ before the end of the evaporation to provide strong vertical confinement, with trapping frequency $\omega_z = 2\pi \times 604(10)$ Hz for $1/k_Fa<0.5$ and $\omega_z = 2\pi \times 480(10)$ Hz otherwise. Successively, in the $x$–$y$ plane, a repulsive cylindrically symmetric potential is turned on to trap the resulting sample in a circular region. This circular box is projected using a Digital Micromirror Device (DMD). When both potentials have reached their final configurations, the infrared lasers forming the crossed dipole trap are adiabatically turned off, completing the transfer into the final pancake trap. We set the internal $R_i = \SI{20\pm0.5}{\mu m}$ and external $R_o = \SI{30\pm0.5}{\mu m}$ radii for $1/k_F a > 0.5$, otherwise, $R_i = \SI{18.7\pm0.5}{\mu m}$ and $R_o = \SI{25\pm0.5}{\mu m}$. These two configurations were chosen as a compromise between reducing the ring width and preserving stable circulation states. All over the main text, we refer to the radius as the mean value in our small-width ring, calculated as $\bar R = (R_i + R_o)/2$. A residual radial harmonic potential of $\SI{2.5}{Hz}$ is present due to the combined effect of an anti-confinement provided by the $\mathrm{TEM_{01}}$ laser beam in the horizontal plane and the confining curvature of the magnetic field used to tune the Feshbach field. This weak confinement has a negligible effect on the sample over the $R_o=\SI{30 \pm 0.5}{\mu m}$ radius of our box trap, resulting in an nearly homogeneous density (see Ext.~Fig.~\ref{extfigMu}a). We estimate the Fermi energy as \cite{hernandez2024connecting}:
\begin{equation}
    \frac{E_F}{h}=\left(\frac{\hbar  \omega_z N_p }{m \pi^2 (R_o^2 - R_i^2)} \right)^{1/2},
\end{equation}
and the chemical potential of the unitary Fermi gas is:
\begin{equation}\label{eq:muufg}
    \mu= \xi^{3/4} h \left(\frac{\hbar  \omega_z N_{\rm p} }{m \pi^2 (R_o^2 - R_i^2)} \right)^{1/2},
\end{equation}
where $\xi$ is the Bertsch parameter. In the BEC regimes, the chemical potential is estimated to be:
\begin{equation}\label{eq:mubec}
    \mu_M = \left( \frac{3}{2} \frac{\hbar^2 \omega_z N_p a_M}{\sqrt{m} (R_o^2-R_i^2)}\right)^{2/3},
\end{equation}
where $a_M =0.6\,a$ corresponds to the molecule-molecule scattering length. Across the BEC–BCS crossover, we remain in the three-dimensional thermodynamic regime, i.e., $\mu\gg \hbar\omega_z$. As shown in Ext.~Fig.~\ref{extfigMu}, the rate $\mu/\hbar\omega_z>4$ for $1/k_Fa<2$ while $\mu/\hbar\omega_z\sim2$ otherwise, ensuring that we never enter the full-2D regime. However, some effects related to the dimensional crossover are expected to manifest for $1/k_Fa>2$. The Fermi energy, instead, always remains around $20\, \hbar \omega_z$ irrespective of the interaction regime.

After preparing our ring geometry with homogeneous planar density, we exploit the DMD to produce a phase imprinting in order to excite persistent current with well-defined circulation state. As shown in Ref.[~\citenum{DelPace2022}], the phase imprinting is performed by shining the superfluid ring with an optical azimuthal gradient for a time shorter than the characteristic time for density response. The light imprints a total phase $\Delta \phi_{I}$ proportional to the controllable light pulse duration. As result, this protocol sets high-fidelity winding number $\textit{w}$, obtained from time-of-flight interferometric measurements, that showcase a stair-like behavior as function of $\Delta \phi_{I}$. In the presence of this circulation, we apply 100 ms after a sinusoidal perturbation of the form $V_\text{pert} (\theta) = -V_0 \cos{\theta}$. We keep the perturbation potential for 50 ms in order to allow the atomic system to equilibrate and dampen possible spurious excitations. Finally, we quench off the perturbation and start studying the dynamics of the system by probing the atoms with \textit{in-situ} absorption imaging. 

From the \textit{in-situ} images of the atomic superfluid, we extract the normalized azimuthal density profile integrating the planar density $n_{\text{2D}}(r,\theta) = \int n_{\text{3D}}(r,\theta, z) \ dz$ over the radius: $n_\theta (\theta) = \int n_{\text{2D}}(r,\theta) \ r \ dr / N$.  Ext.~Fig.~\ref{extfigMu}b-c show the 2D density of the unwrapped ring and its corresponding normalized azimuthal density profile $n_\theta (\theta)$, which gives the fraction of atoms per unit angle. Here, the density is modulated due to the presence of the perturbation $V_\text{pert}$. We can estimate the height of this applied perturbation by using the local density approximation: $\mu_\text{loc} (\theta) = \mu - V(\theta)$, and the polytropic approximation: $\mu = g_{\gamma} n^\gamma$, where $g_{\gamma}$ is a constant and $\gamma$ is the polytropic exponent. Combining these two approximations results in $n_\theta (\theta) = \left[ (\mu - V_\text{pert} (\theta))/g_{\gamma} \right]^{1/\gamma}$. Moreover, we can consider the limiting conditions $n_\theta (\theta=0) = n_\text{max} = \left[ (\mu + V_0)/g_{\gamma} \right]^{1/\gamma}$ and $n_\theta (\theta=\pi) = n_\text{min} = \left[ (\mu - V_0)/g_{\gamma} \right]^{1/\gamma}$, and extract:
\begin{equation}
    \frac{V_0}{\mu} = \frac{\left(\frac{n_\text{max}}{n_\text{min}}\right)^\gamma - 1}{\left(\frac{n_\text{max}}{n_\text{min}}\right)^\gamma + 1} \ .
\end{equation}
Here, the minimum and maximum densities can be extracted from a sinusoidal fit of the density profile $n_\theta (\theta)$. The polytropic exponent takes the values $\gamma = 1$ for the BEC regime and $\gamma = 2/3$ for the unitary Fermi gas (UFG) and BCS regimes. For all conditions explored in this work, we used a perturbation height of $V_0/\mu \lesssim 0.1$, which is typically a safe threshold for a linear response of the system.

\subsection*{Fitting procedure}

\begin{figure}[b!]
\centering
\includegraphics[width=8.6cm]{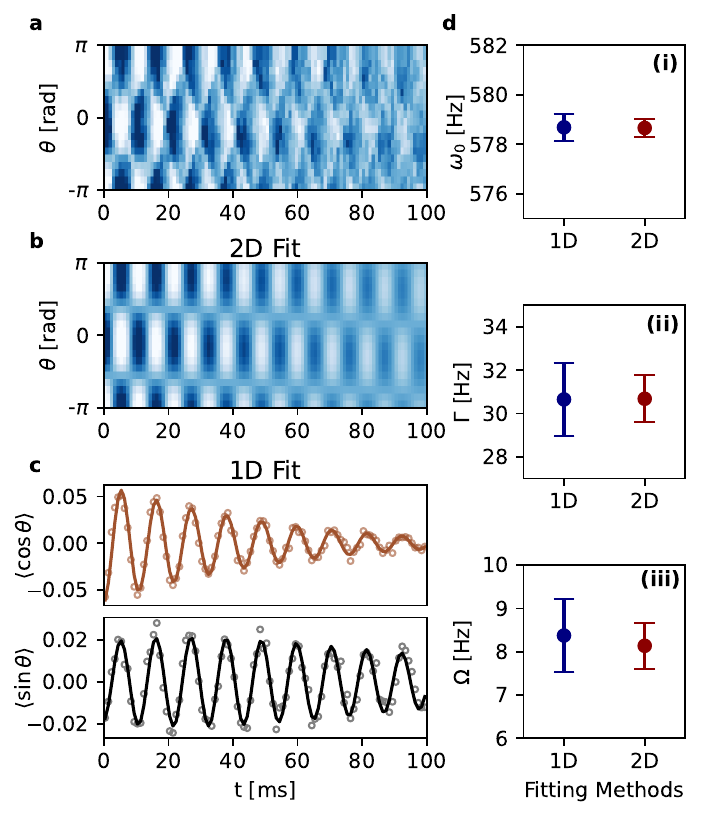}
\caption{\textbf{Fitting procedures.} 
(\textbf{a}), Temporal evolution of the normalized azimuthal density in a rotating superfluid of winding number $\textit{w}=1$, in the unitary regime. (\textbf{b}), 2D fit applied to the interferogram shown in a. (\textbf{c}), calculated values of $\langle \cos{\theta} \rangle$ and $\langle \sin{\theta} \rangle$, with their corresponding 1D sinusoidal fit (brown and black curves, respectively). (\textbf{d}), results for the phonon frequency (\textbf{i}), the decay rate (\textbf{ii}) and the precession frequency (\textbf{iii}) extracted with the two fitting method.
}
\label{fig_SM_fitting}
\end{figure}

The applied small-amplitude perturbation potential $V_{\text{pert}}$ on the superfluid ring excites two counter-propagating phonon modes. As a consequence, the density of the system will be given by the combination of the initial homogeneous density $n_{\theta}^0$ and the two contributions $\delta n_\pm$ for the clockwise and counter-clockwise propagating excitations:
\begin{equation}
    n_{\theta}(\theta, t) = n_{\theta}^0 + \delta n_+(\theta, t) + \delta n_-(\theta, t) = n_{\theta}^0 +\delta n (\theta, t)
\label{azimuthal_density}
\end{equation}
Here, the total density modulation $\delta n (\theta, t)$ is
\begin{align}\label{eqfit2d}
\begin{split}
    \delta n (\theta, t) = \frac{1}{2} V_0 e^{- \frac{\Gamma t}{2}} &\left[\chi_+\cos(\theta-\theta_0-\omega_+t)+\right.\\
    &\left.\chi_-\cos(\theta-\theta_0+\omega_-t)\right]
\end{split}
\end{align}
Alternatively, 
\begin{align}
\begin{split}
    \delta n (\theta, t) = \frac{V_0\chi}{2} e^{- \frac{\Gamma t}{2}} &[\cos(\theta-\Omega t)\cos(\omega_0t)+\\
    &\frac{2\Delta_c}{c_s}\sin(\theta-\Omega t)\sin(\omega_0t)]
\end{split}
\end{align}
where a phenomenological damping $\Gamma$ was added to account for any decay mechanism of the excitation, $\theta_0$ is an angular offset, and $\omega_{\pm} = \omega_0\pm \Omega$ are the Doppler shifted phonon frequencies, where we recall the identity $\Omega=\Delta_c / \bar R$ relating the rotation of the system $\Omega$, to the Doppler shift velocity $\Delta_c$. Eq.\eqref{eqfit2d} provides a two-dimensional fit we perform on the interferograms $\delta n (\theta, t)$, allowing to extract the parameters $\omega_0$, $\Gamma$ and $\Omega$.

Another way to study the evolution of the azimuthal density consists in calculating the density-weighted angular moments $\langle \cos{\theta} \rangle (t) = \int n_{\theta}(\theta, t) \cos \theta\, d\theta$ and $\langle \sin{\theta} \rangle (t) = \int n_{\theta}(\theta, t) \sin \theta\, d\theta$. This set of observables, are equivalent to the temporal evolution of the angular Fourier decomposition of the density profile. In particular, Eq. \eqref{eqfit2d} can be reformulated as:
\begin{align}
\begin{split}
    s (t) &= \iint n_{2D}(\vec{r},t) e^{-i\theta} d\vec{r} = \langle \cos \theta \rangle + i\langle \sin \theta \rangle,\\
    s (t) &= \frac{\pi V_0}{2}  e^{-\Gamma t/2} \left(\chi_-e^{-i \omega_0t} + \chi_+ e^{i \omega_0t} \right)e^{i (\theta_0+\Omega t)}.\label{eqsincos}    
\end{split}
\end{align}
Similarly to the interferogram $\delta n (\theta, t)$, we perform the fitting procedure of $\langle \cos{\theta} \rangle (t)$ and $\langle \sin{\theta} \rangle (t)$ simultaneous using Eq.\eqref{eqsincos}.

Extended Fig.~\ref{fig_SM_fitting} shows an example of the interferogram in an unitary Fermi superfluid. Additionally, we show the fitting results from Eq.~\eqref{eqfit2d} labeled "2D Fit", and Eq.~\eqref{eqsincos} labeled as "1D Fit". The two fitting procedures provide compatible results, as shown in panel d. In general, we found the fit method using Eq.~\eqref{eqsincos} to be more resilient against noise, particularly when the amplitude of the perturbation is very small. For this reason, in the main text we report the results from the fitting procedure from Eq.~\eqref{eqsincos}.

\subsection*{Interferograms in BCS superfluids}
\begin{figure}[b!]
\centering
\includegraphics[width=8.6cm]{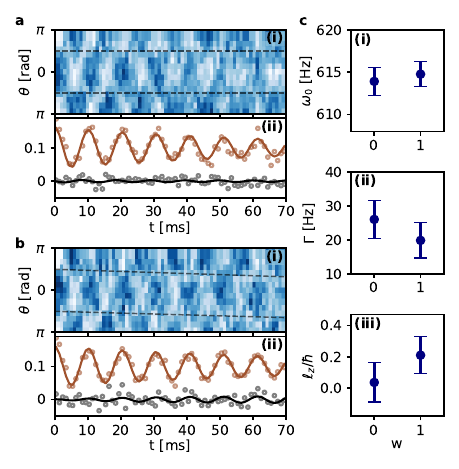}
\caption{\textbf{BCS interferograms.} Temporal evolution of the normalized azimuthal density in the absence of circulation $\textit{w}=0$ (\textbf{a}) and in a rotating superfluid of winding number $\textit{w}=1$ (\textbf{b}), both in the BCS regime with $1/k_Fa = -0.55(1)$. Panels \textbf{a}(ii) and \textbf{b}(ii) show the calculated values of $\langle \cos{\theta} \rangle$ (brown) and $\langle \sin{\theta} \rangle$ (black), with their corresponding 1D sinusoidal fits. (\textbf{c}), Results extracted for each winding number, where we can see the phonon frequency (\textbf{i}), the decay rate (\textbf{ii}) and the precession frequency (\textbf{iii}).}
\label{fig_SM_BCS}
\end{figure}

To extend the analysis presented in Fig.~2 and Fig.~3, we include the interferometric data, angular moment evolution, and fitting results for the BCS superfluid at $1/k_Fa = -0.55(1)$ (see Ext.~Fig.~\ref{fig_SM_BCS}). Consistent with the observations in the BEC and unitary regimes, the interferograms reveal a slow drift of the standing-wave pattern, which is quantitatively captured by the dynamics of the density-weighted angular moments.

\subsection*{Speed of sound propagation: geometrical factor}

To obtain the speed of sound across the BEC–BCS crossover, we employ the polytropic approximation $\mu\propto n^{\gamma}$. In the BEC $\gamma=1$, while in the UFG $\gamma=2/3$. Following Ref.[\citenum{joseph2007measurement}], the wave front speed is given by
\begin{equation}
    \bar{c}_s = \sqrt{\frac{1}{m}\frac{\int n r dr dz}{\int\left(\frac{\partial \mu_\text{loc}(z)}{\partial n}\right)^ {-1} r dr dz}},
\end{equation}
where the subscript ``$\text{loc}$'' indicates the local chemical potential obtained by assuming a local density approximation: $\mu_\text{loc}(z) = \mu - V(z)$. Writing $\mu = g_{\gamma}n^{\gamma}$, with $g_{\gamma}$ a constant, the wave front speed is:
\begin{align}
    m\bar{c}_s^2 &= \gamma g_{\gamma}\frac{\int n(z)dz}{\int n^{1-\gamma}(z) dz} = m A_{z}^2 c_{0}^2
\end{align}

Where $c_0$ correspond to the peak speed of sound at the center of the trap, and $A_{z}$ a geometrical correction factor:
\begin{equation}
    A_{z} = \sqrt{\frac{\Gamma\left(1 + \frac{1}{\gamma}\right)\Gamma\left( \frac{3}{2}+ \frac{1-\gamma}{\gamma}\right)}{\Gamma\left( \frac{3}{2}+ \frac{1}{\gamma}\right)\Gamma\left(1 + \frac{1-\gamma}{\gamma}\right)}}
\end{equation}
where $\Gamma$ is the Gamma function.

In addition to the effect of the vertical confinement, the sound also propagates along different radii from $R_i$ to $R_o$, hence a correction factor should also be considered. 
The angular frequency $\omega_0 = \bar{c}_s/r$, so the radial average phonon frequency, $\langle \omega_0\rangle_r$, is given by:
\begin{align}
    \langle \omega_0\rangle_r &=  \frac{1}{R_o-R_i} \int_{R_i}^{R_o} \frac{\bar{c}_s}{r} dr  = \frac{\log(R_o/R_i)}{R_o-R_i} \bar{c}_s
    \\
    \langle \omega_0\rangle_r &= \frac{\log(R_o/R_i)(1+R_o/R_i)}{2(R_o/R_i-1)} \bar{\omega}_0 = A_r \bar{\omega}_0
\end{align}
where $\bar{\omega}_0 = \bar{c}_s/\bar R$, and $A_r = \frac{1}{2}\log(R_o/R_i)(R_o+R_i)/(R_o-R_i)$. 
The excitation protocol allows us to experimentally determine $\langle \omega_0\rangle_r$. In order to recover the behavior of $c_s/v_F$ for homogeneous systems, we therefore need to consider the correction terms $A_z$ and $A_r$. 

\begin{equation}
    c_0 = \frac{\bar{c}_s}{A_z} = \frac{\bar R \bar{\omega}_0}{A_z} = \frac{\bar R }{A_z A_r}\langle \omega_0\rangle_r.
\end{equation}

In particular, for the data at unitarity, the geometrical coefficients are: $A_z=\sqrt{3/4}\sim0.866$, and $A_r\sim1.00689$. To estimate the Bertsch parameter at unitarity, let us recall the expression linking the speed of sound to the Fermi velocity: $c_s^2/v_F^2 = \xi_B/3$. For the correct determination of $\xi_B$, we should consider the geometrical factor $A_z A_r$ at unitarity. The value of $\xi_B$ reported in the main text corresponds to the weighted mean over 10 different experimental realizations, reporting the standard error of the weighted mean as error. We emphasize that the largest source of error originates from the stability in the number of atoms $\delta N/N$ over the full experimental run. Additionally, the small size of our ring makes our experimental radial resolution, $\delta R/R\sim3\%$. Employing larger systems would improve the estimation of $\xi_B$.

\subsection*{Linear response of the system}

The velocity of sound propagating in a uniform superfluid moving with velocity $v$  exhibits a Doppler shift fixed by the law $c_\pm = c_0 \pm v$, with $c_0$ the velocity of sound in the absence of permanent current.  The above result follows from a simple Galilean transformation to the frame where the system is at rest. It is applicable to finite temperature only if the superfluid and normal components of the system move with the same velocity. In the present work, we are instead considering the case where the circulating current is given only by the superfluid component, while the thermal component remain at rest. As a consequence the two sound velocities predicted by Landau's hydrodynamic theory of superfluids (first and second sound) will exhibit different (anomalous) Doppler shifts, as first pointed out in liquid He-II and recently considered in the case of a supersolid dipolar gas \cite{Tomasz2025}. Useful insight into the problem can be obtained by investigating the behavior of the linear density response function. 

Let us first consider the case where both the superfluid and normal components are at rest. In this case, the density response function, for small wave vectors $k$ and frequencies $\omega$, is characterized by the presence of two poles for positive values of $\omega$, associated with the excitation of first and second sound in the collisional hydrodynamic regime \cite{Hohenberg1965,Hu2010}:
\begin{align}
\begin{split}
    \chi^{\text{two sounds}}(k,\omega) &=-N\frac{k^2}{m}\left[\frac{Z_1}{\omega^2-c_1^2k^2} +\frac{Z_2}{\omega^2-c_2^2k^2} \right],\\
&= -N\frac{k^2}{m}\frac{\omega^2-\gamma c^2_{2,0}k^2}{(\omega^2-c^2_1k^2)(\omega^2-c^2_2k^2)} \;,
\end{split}
\label{2soundsNodoppler}
\end{align}
where $N$ is the total number of atoms and, for simplicity, we have ignored collisional damping effects. In eq.(\ref{2soundsNodoppler}) we employ the formalism of uniform systems where the excitation operator corresponds to the Fourier transform $\rho_k= \sum_j \exp(ikx_j)$ of the density operator. In the case of the ring geometry with radius $\bar R$, the wave vector $k$ is oriented along the azimuthal direction $\theta $ and the excitation operator takes the form $\rho=\sum_j \exp(ik \bar R\theta_j)$ where $k$ is quantized according to the rule $k=\textit{w}/\bar R$, with $\textit{w}=0, \pm 1, \pm 2 ..$, ensuring the proper periodicity condition. 

In the above equation $c_1$ and $c_2$ are the first and sound velocities predicted by two fluid Landau's hydrodynamic theory and we have introduced the isentropic expansion coefficient 
\begin{equation}    
\gamma = \frac{(\partial p/\partial \rho)_T}{(\partial p/\partial \rho)_p}=\frac{c^2_s}{c^2_T} \; ,
\label{gamma}
\end{equation}
defined by the ratio between the inverse isothermal and the adiabatic compressibilities, while  $c_T=\sqrt{(\partial p/\partial \rho)_T}$ and $c_s=\sqrt{(\partial p/\partial \rho)_s}$  are the corresponding isothermal and isentropic velocities. The quantity $c_{2,0}=\sqrt{(s^2/m)(\rho_s/\rho_n)(\partial T/\partial s)_p}$ is the uncoupled value of the second sound velocity, corresponding to the actual value of $c_2$ in the limit of negligible thermal expansion, i.e if $\gamma = 1 $. In this same limit, the first sound velocity approaches the isentropic sound velocity $c_s$. 

The weights $Z_1$  and $Z_2$ entering Eq.(\ref{2soundsNodoppler}) correspond to the relative contributions of the two modes to the energy weighted f-sum rule $Nk^2/m$ and satisfy the condition $Z_1+Z_2=1$ which follows from Galilean invariance. Important quantities are also the contribution of the two sounds to the inverse energy weighted moment of the dynamic structure factor, given by $\chi_1=Z_1/c_1^2$ and $\chi_2=Z_2/c_2^2$, respectively. The sum of the two terms corresponds to the static response $\chi\equiv \chi(k\to 0, \omega=0) $, which, according to the compressibility sum rule, coincides with the isothermal compressibility $(\partial \rho/\partial p)_T$:
\begin{equation}
\chi_1+\chi_2=\chi\equiv(\partial \rho/\partial p)_T \; .
\label{kappaT}
\end{equation}
By explicitly calculating the poles and the residues of the response function (\ref{2soundsNodoppler}), one finds that the ratio $\chi_2/\chi_1$ can be conveniently expressed in the form \cite{Vinen1971,Hu2010}
\begin{equation}
    \frac{\chi_2}{\chi_1} = \frac{1}{\gamma -1}\left(c_1^2/c^2_T-1\right)^2 \; .
    \label{Vinen}
    \end{equation}
By using the expansion \cite{Hu2010}
\begin{equation}
c^2_1= c^2_s(1 + (\gamma-1)c^2_2/c^2_s+..)
\label{c1cs}
\end{equation}
one finds that in weakly compressible systems, where $\gamma\simeq 1$, the quantity $c_1^2/c^2_T$ approaches the thermal expansion coefficient $\gamma$. The ratio $\chi_2/\chi_1$ then becomes proportional to the Landau Placzek ratio $\epsilon_{LP}= \gamma -1$, according to:
\begin{equation}
    \frac{\chi_2}{\chi_1} = (\gamma-1)(1+2c^2_2/c^2_1 +..)  
    \label{W21expansion}
    \end{equation}
and is consequently negligible, unless one considers regimes close to the critical temperature \cite{Patel2020}.

\subsection*{Density response in presence of a permanent current }
In this section we describe the behavior of the density response function in the presence of a permanent current, assuming that the ratio $\chi_2/\chi_1$ between the contributions of second and first sound to the static response function of a Fermi superfluid can be safely ignored. As discussed in the previous section this is an accurate approximation not only at zero temperature, where only one sound (the Anderson Bogoliubov sound) propagates, but also at higher temperatures in the case of the weakly compressible unitary Fermi gas where the contribution of second sound to the density response function is very small.

The presence of a permanent current flowing in the system causes the Doppler shift of the sound wave in two components propagating parallel or anti-parallel to the current, yielding the following form for the response function
\begin{equation}
\chi^{\text{one sound}}(k,\omega) =- N\frac{k^2}{m}\frac{1}{(\omega-\omega_+)(\omega+\omega_-)} \; ,
\label{onesound}
\end{equation}
where, for simplicity, we have omitted the symbol $1$ of first sound and introduced the frequencies $\omega_\pm$  of the two phonon components  propagating parallel or antiparallel to the stationary current. These poles are characterized by the dispersion relation $\omega_\pm=kc_s(1 \pm \Delta_c/c_s)$ where $\pm \Delta_c$ are the corresponding Doppler shifts.

The large $\omega$ expansion of the density response function provides the result:
\begin{equation}
\chi^{\text{one sound}}(k,\omega)_{\omega \to\infty}= -\frac{1}{\omega^2}N\frac{k^2}{m}-\frac{1}{\omega^3} N\frac{k^3}{m}2\Delta_c -..
\label{expansion}
\end{equation}
which should be compared with the general result for the expansion of the response function predicted by many-body theory \cite{PitaevskiiStringariBook2016}:
\begin{equation}
 \chi(k,\omega) _{\omega \to\infty}= -\frac{1}{\omega^2}m_1^+- \frac{1}{\omega^3}m_2^- -...
\label{expansion*}
\end{equation}   
where 
\begin{align}
\begin{split}
m_1^+&=
\int (d\omega) \omega (S(k,\omega)+S(-k,\omega))= \langle [\rho_k,[H,\rho_{-k}]]\rangle \\ 
&=N\frac{k^2}{m}
\end{split}
\label{m1}
\end{align} 
is the model independent f-sum rule, given by the sum of the energy weighted moments of the dynamic structure factors $S(k,\omega)$ and $S(-k,\omega)$, 
while
\begin{align}
\begin{split}
m_2^-&=\int d\omega \omega^2 (S(k,\omega)-S(-k,\omega))= \langle[[\rho_k,H],[H,\rho_{-k}]]\rangle \\
&=N\frac{k^3}{m^2}2\langle P_x\rangle  
\end{split}
\label{m2}
\end{align}
is the difference between the corresponding twice energy weighted moments, fixed by the momentum per particle $\langle P \rangle$ associated with the permanent current. In the case of a ring, the momentum $P_x$ should be replaced by the momentum $P_\theta$ oriented along the azimuthal angle $\theta$ and the angular momentum per particle is accordingly given by $\ell_z= \bar  R \langle P_\theta\rangle$. It is worth noticing here that in Galilean invariant systems, satisfying the condition $[H,P_x]=0$, both moments $m_1^+$ and $m_2^-$ are exhausted by the contribution of the gapless phonons, gapped states as well as multiphonon excitations providing contributions of order $k^4$, hence negligible in the  limit of small wave vectors $k$. From the comparison between Eqs. (\ref{expansion}) and (\ref{expansion*}) one then concludes that on one hand phonons exhaust, as expected,  the f-sum rule (\ref{m1}) and on the other hand, comparing the terms in $1/\omega^3$ in the expansion, one finds that  the Doppler shift $\Delta_c$ is directly related to the momentum $\langle P_x \rangle$ per particle associated with the permanent current present in the system:
\begin{equation}
\Delta_c=\frac{1}{m}\langle P_x\rangle=f_sv_s+f_nv_n \; ,\label{fsfn}
\end{equation}
where $v_s$ and $v_n$ are the permanent velocities of the superfluid and of the normal components, while $f_s$ and $f_n$ are the fractions of the superfluid and normal fluids in the system, satisfying the normalization condition $f_s+f_n=1$. At zero temperature, where $f_n=0$ and $f_s=1$,
the Doppler shift coincides with  the value of the superfluid velocity $v_s$. At finite temperature the situation is different. In particular, if only the superfluid is moving,  the Doppler shift gives direct access to the quantity $f_s v_s$ and is consequently  explicitly sensitive to the superfluid fraction present in the system. 

The energy weighted and the twice energy weighted moments, Eqs. (\ref{m1}) and (\ref{m2}), calculated with the different choice $(x\pm iy)^2$ for the excitation operators,  yielding $m_1^+= 8(\hbar^2/m)N\langle x^2+y^2\rangle$ and $m_2^-=16(\hbar^3/m^2)N\ell_z$, were used in Ref.[~\citenum{zambelli1998}] to calculate the Doppler splitting $\omega_+-\omega_-=(2/m)\ell_z/\langle x^2+y^2\rangle$ exhibited by the quadrupole collective mode  in terms  of the angular momentum per particle $\ell_z$ carried by a quantized vortex in a harmonically trapped quantum gas. The resulting precession of the quadrupole shape was then employed in Ref.[~\citenum{chevy2000measurement}] to infer the value of $\ell_z$ and the corresponding value of the quantum of circulation $\kappa=h/m$ in a Bose-Einstein condensed atomic gas.

\begin{figure}[t!]
\centering
\includegraphics{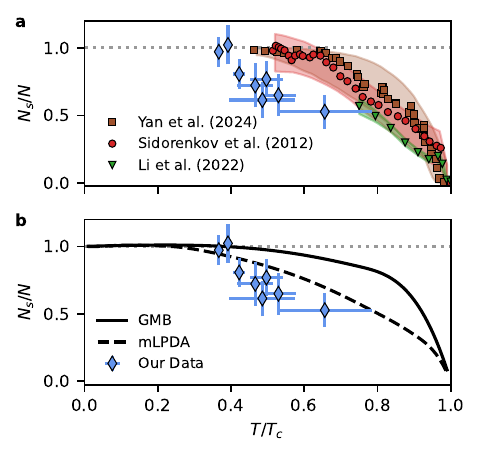}
\caption{\textbf{Superfluid fraction at unitarity.} 
(\textbf{a}), Comparison between our results (diamonds) with previous measurements of the superfluid fraction for homogeneous systems. Data from Ref.[~\citenum{sidorenkov2013second}] is shown in red, data from Ref.[~\citenum{yan2024thermography}] is shown in green, and data from Ref.[~\citenum{Li2022}] in brown. (\textbf{b}), Comparison between our results (diamonds) with theoretical models of the superfluid fraction for homogeneous systems. In particular, modified Local Phase Density Approximation mean-field description ~\cite{pisani2023inclusion} (black dashed curve), and the extended GMB approach~\cite{pisani2023inclusion, pisani2018gap} (black continuous). 
}
\label{fig_SM_supfraction}
\end{figure}

\subsection*{Superfluid fraction at unitarity}

\subsubsection*{Comparison 3D homogeneous data}
In Ext.~Fig.~\ref{fig_SM_supfraction}, we report the comparison between our experimental estimation of the superfluid fraction to the experimental values reported in Refs.[~\citenum{yan2024thermography, Li2022, sidorenkov2013second}], as well as two distinct theoretical models from Ref.[~\citenum{pisani2023inclusion}]. As shown, our estimation are significantly lower than those reported experimentally for homogeneous system. A similar fact can be said about the comparison with the extended Gorkov–Melik–Barkhudarov (GMB) approach of Ref.[~\citenum{pisani2023inclusion, pisani2018gap}]. 

\bigskip
\subsubsection*{Trap average superfluid fraction}
To estimate the superfluid fraction at unitarity in our trap geometry, we exploit the fact that the superfluid density $\rho_s/\rho$ is only a function of the reduced temperature $T/T_c$, and that in our system we can employ the local density approximation. We rely on the equation of state (EOS) reported in Ref. [\citenum{ku2012revealing}], where the density of a unitary Fermi gas was written in the local density approximation:
\begin{equation}
n_{3D}(\vec{r};\mu,T) = \frac{1}{\lambda_{T}^3} f_n\left[\beta(\mu-V(\vec{r}))\right],
\end{equation}
where $\beta=1/(k_B T)$ and $\lambda_{T}=\sqrt{2\pi \hbar^2/ mk_B T}$ is the thermal de Broglie wavelength, $m$ the mass of a $^6$Li atom, $V(\vec{r}) = \frac{1}{2} m\omega_z^2 z^2 + U_{box}(r, \theta)$ the confining potential, and the equation of state $f_n(q)$ defined as \cite{Del_Pace2021}:
\begin{equation}
f_n(q) = \left\{
     \begin{array}{lr}
       \sum_{k=1}^4  k\ b_k\ e^{kq}& q<-0.9\\
       -\mathrm{Li}_{3/2}(-e^q)F(q) & -0.9<q< 3.9 \\
       \frac{4}{3\sqrt{\pi}} \left[\left(\frac{q}{\xi}\right)^{\frac{3}{2}} - \frac{\pi^4}{480} \left(\frac{3}{q}\right)^{\frac{5}{2}}\right]& q>3.9\\
     \end{array}
   \right.
\end{equation}
where $F(q)=n(q)/n_0(q)$ is the ratio between the unitary and the non-interacting Fermi gas density measured experimentally in Ref.[~\citenum{ku2012revealing}] and Li$_{3/2}(x)$ is the polylogarithm of order $3/2$ and argument $x$. To obtain the superfluid density, we exploit the fact that $f_s^{3D}(T/T_c) =\rho_s(T/T_c)/\rho$ is only a function of the reduced temperature $T/T_c$ and consider as superfluid density:
\begin{equation}
n_{s}(\vec{r};\mu,T) = \frac{1}{\lambda_{T}^3} f_s^{3D}\left(\frac{k_BT}{\epsilon_F}\right) f_n\left[\beta(\mu-V(\vec{r}))\right],
\end{equation}
where $\epsilon_F = \epsilon_F^0 (n/n_0)^{2/3}$, where the index '0' corresponds to the values of density and Fermi energy at the center of the trap. The trap-averaged superfluid fraction can therefore be written as:
\begin{equation}\label{avgsupfrac}
    \frac{N_s}{N} = \frac{\int f_s^{3D}\left(\frac{T}{T_F} \frac{T_F}{T_c}\right) f_n\left[\beta(\mu-V(z))\right] dz}{\int f_n\left[\beta(\mu-V(z))\right] dz}.
\end{equation}
where $T_c/T_F=0.17$. To compute the estimations of Fig.~\ref{fig4} of the main manuscript we perform a linear interpolation of the raw data from each data set to evaluate the function $f_s^{3D}(T/T_c)$ in a continuous fashion.

\end{document}